\documentclass[12pt]{article}
\usepackage{amsmath,amssymb,bm,graphicx}
\usepackage{color} 

\newcommand{\tp}{\text{p}}

\newcommand{\sgn}{\text{sgn}}

\newcommand{\da}{{\downarrow }}
\newcommand{\ua}{{\uparrow }}
\numberwithin{equation}{section}
\usepackage{hyperref}
\usepackage{cite}

\begin{document}

%\vskip 0.5 truecm

\begin{center}
{\Large{\bf Seesaw, coherence and neutrino oscillations}}
\end{center}\vskip .5 truecm
\begin{center}
{\bf \large{ Tomi Kupiainen and Anca Tureanu}}

\vspace*{0.4cm} 
{\it {Department of Physics, University of Helsinki, P.O.Box 64, 
\\FIN-00014 Helsinki,
Finland
}}
\end{center}

\begin{abstract}
We present a prescription for consistently constructing non-Fock coherent flavour neutrino states within the framework of the seesaw mechanism, and establish that the physical vacuum of massive neutrinos is a condensate of Standard Model massless neutrino states. The coherent states, involving a finite number of massive states, are derived by constructing their creation operator. Such a construction is the key requirement so that the oscillations of particles indeed occur. We comment on the inherent non-unitarity of the oscillation probability induced by the requirement of coherence. 
\end{abstract}

\vskip 1cm
\begin{flushright}
{\it Dedicated to the memory of Samoil Bilenky}
\end{flushright}

\vskip 1cm

\section{Introduction}\label{intro}

The nature of neutrinos, either as Dirac or Majorana particles \cite{Bilenky:2020wjn}, is directly related to their mass-generating scheme. The seesaw mechanism \cite{minkowski, yanagida, GMRS, Glashow, mohapatra,foot, ma}, with its type-variations, leads up to Majorana neutrinos and is generally considered as one of the most plausible explanations for the small neutrino masses, though the confirmation of the Majorana nature by neutrinoless double beta decay \cite{Schechter-Valle} is still far from being established.
From the point of view of quantum field theory, the question "how do neutrinos oscillate" is at least as intriguing as the questions "how do neutrinos acquire such tiny masses compared to other leptons". Most of the present-day efforts are put into giving an answer to the latter question. The seesaw mechanism ascribes the smallness of the observed neutrino masses to the largeness of a new physics energy scale which controls the mass of neutrinos. In all its versions, it represents concrete models of Weinberg's lepton number violating higher order operators involving the Standard Model Higgs field \cite{weinberg} and usually respecting the Standard Model gauge symmetry. In a more ample context, the seesaw mechanism leads to an explanation of the matter-antimatter asymmetry in the Universe, through leptogenesis \cite{Fukugita-Yanagida-lepto}.
The strong appeal of the seesaw mechanism is partially curtailed by the UV sensitivity of the Higgs mass squared, which is quadratically sensitive to the seesaw scale, thus inducing a fine-tuning problem (see, for example, the review \cite{xing}).

In this paper, our focus will be on the question of "how do neutrinos oscillate", and more specifically, on the coherence of the superposition of massive neutrino states engendered by the seesaw mechanism.
Pontecorvo's extension of the state mixing and oscillation paradigm from the $K_0-\bar K_0$ system \cite{GM-P, Pais-Piccioni} to neutrinos \cite{Pontecorvo, Gribov-Pontecorvo, Bilenky-Pontecorvo} (see also \cite{BP_review}) is based on two {\it sine qua non} requirements: i) the massive neutrino states have different masses; ii) the superposition of the massive neutrino states is coherent. Here, by coherence is meant that the relative phases between the particle states composing a flavour neutrino state are fixed. The second requirement is the source of a deep conceptual problem, since quantum field theory does not give a prescription for defining coherent superpositions of states belonging to different Fock spaces. Our main concern now is to establish such a prescription for the specific case of oscillating neutrino states that are superpositions of Majorana states generated by the seesaw mechanism, using the general formulation proposed in \cite{AT_neutron, AT_neutrino}.

According to the Gribov--Pontecorvo conjecture \cite{Gribov-Pontecorvo}, the flavour neutrino states are defined as
\begin{eqnarray}\label{state_mix_N}
|\nu_{l}\rangle=\sum_{i=1}^N U^*_{li} |\nu_{i}\rangle, \ \ \ \ \ l=e,\mu,\tau,
\end{eqnarray}
where $|\nu_{i}\rangle,\  i=1,2,\ldots, N$ are massive (Majorana) neutrino states, $U$ is the PMNS mixing matrix of neutrino fields, and $L$ and $R$ denote the left and right chiralities. The latter appears in the diagonalization of the seesaw Lagrangian by the change of variables from the flavour fields $\nu_{lL}(x)$  and the sterile fields $\nu_{sR}(x)$ to the massive fields $\nu_{iL}(x)$:
\begin{eqnarray}\label{field_mix_N}
\nu_{lL}(x)=\sum_{i=1}^N U_{li} \nu_{iL}(x),\quad\quad C\bar\nu_{sR}^T(x)=\sum_{i=1}^N U_{si} \nu_{iL}(x),
\end{eqnarray}
where $C$ is the charge conjugation matrix.
To this definition, one has to add the {\it supplementary assumption of coherent superposition} of the massive neutrino states in \eqref{state_mix_N}. This assumption is as old as the first proposal of particle mixing in the $K_0-\bar K_0$ system -- the coherence of superposed particle states is obtained by default if the states have the same mass (i.e. belong to the same Fock space) \cite{GM-P, Pais-Piccioni}, but not when the states have different masses. There is a rich literature regarding various definitions of flavour states in quantum field theory, including also coherence (for a far-from-complete selection, see \cite{Chang} -- \cite{KF_PI}). Here, we shall pursue another approach, generalizing the technique previously proposed for the construction of coherent oscillating particle states in the context of neutron-antineutron \cite{AT_neutron} and Dirac neutrino oscillations \cite{AT_neutrino}. 

In this paper we show that the seesaw oscillating neutrino states can be obtained as {\it intrinsically coherent superpositions} of massive neutrino states expressed by the formula
\begin{eqnarray}\label{coh_state_mix_N}
|\nu_{l} ({\bf p})\rangle=\sum_{i=1}^N U^*_{li} \sqrt{\frac{ E_{i\tp}+\tp}{2 E_{i\tp}}}|\nu_{i}({\bf p})\rangle,
\end{eqnarray}
where $ E_{i\tp}=\sqrt{{\bf p}^2+m_i^2},\ i=1,2,\ldots,N$ are the energies of the neutrinos of mass $m_i$ and momentum ${\bf p}$. By intrinsic coherence it is meant that the feature is (mathematically) built in the states by definition, and does not necessitate supplementary quantum mechanical arguments. 
%Essentially, we find the particle operator which creates the oscillating coherent states \eqref{coh_state_mix_N} out of the vacuum. 

For this construction we make use of the only genuine coherent state \`a la Klauder--Sudarshan--Glauber  \cite{Klauder, Glauber, Sudarshan} that can be built in quantum field theory, namely the vacuum state of a theory with a mass gap \cite{Bog-Shirk}. 
The extra factors are the coefficients of the Bogoliubov transformations which express the interaction-induced mass gap between the Standard Model massless neutrinos and the massive neutrinos of the seesaw mechanism. The oscillating flavour states defined by \eqref{coh_state_mix_N} are not orthogonal, what is a direct consequence of their coherence, since coherent states {\it always} overlap \cite{Klauder, Glauber, Sudarshan}. Nevertheless, for ultrarelativistic neutrinos, the overlap of flavour neutrinos (that would translate experimentally into zero-length conversion of one flavour into another) is way below the present detection accuracy, except for future low-energy neutrino experiments.

The structure of the paper is the following. In Sec. \ref{Lagr_formalism} we collect the basic formulas of the seesaw mechanism \cite{Bilenky,Giunti} for fixing the notation and for further reference; the reader familiar with the subject can skip to the next section. In Sec. \ref{osc_seesaw} we construct the coherent oscillating states for Dirac--Majorana neutrinos, which can straightforwardly be particularized to the type I/III and II seesaw schemes. We show how the coherence is built in the oscillating states, by performing only coherence-preserving transformations. The technique is inspired by Bogoliubov's treatment of superconductivity \cite{Bogoliubov} and the Nambu--Jona-Lasinio scheme \cite{NJL} for the dynamical mass generation of nucleons, adapted here to the case of fields with Majorana mixing terms. The same results are obtained in Appendix \ref{diag} by using the direct procedure of Hamiltonian diagonalization. Furthermore, we discuss also the normalization and the orthogonality of the oscillating neutrino states. In Sec. \ref{outlook} we view the results in a wider context, including the potential effects on the interpretation of the KATRIN and PTOLEMY experiments, in which the neutrinos are non-relativistic.

\section{Lagrangian description of seesaw mechanism }\label{Lagr_formalism}

We consider a mixed seesaw mechanism for one generation of neutrinos, with the Lagrangian:
\begin{eqnarray}\label{Lagr_osc}
{\cal L}&=&\overline{\nu}_L(x)i\gamma^{\mu}\partial_{\mu}\nu_L(x)+\overline{\nu}_R(x)i\gamma^{\mu}\partial_{\mu}\nu_R(x)\cr
& -& m_D\big(\overline{\nu}_L(x)\nu_R(x)+\overline{\nu}_R(x)\nu_L(x)\big)\nonumber\\
&-&\frac{m_L}{2}\big(\nu_L^{T}(x)C\nu_L(x) +\overline{\nu}_L(x)C\overline{\nu}_L^{T}(x)\big)\nonumber\\
&-&\frac{m_R}{2}\big(\nu_R^{T}(x)C\nu_R(x) +\overline{\nu}_R(x)C\overline{\nu}_R^{T}(x)\big),
\end{eqnarray}
where $m_D$, $m_R$, and $m_L$ are real parameters satisfying $m_L\ll m_D\ll m_R$ and $C$ is the charge conjugation matrix (see, for example, \cite{mohapatra-pal,Fukugita,Bilenky,Giunti, valle-romao})\footnote{This choice of real mass parameters implies conservation of $CP$ symmetry. The most general case is when all three masses are complex. However, two of the phases can be absorbed by rephasing the fields $\nu_L$ and $\nu_R$, which leaves only one $CP$ violating Majorana phase. The diagonalization of \eqref{Lagr_osc} with complex $m_L$ can be found, for example, in \cite{Giunti}. For our purposes, it is sufficient to consider the $CP$ even case.}. The field $\nu_L(x)$ is the active neutrino field which appears in charged- and neutral-current weak interactions. The field $\nu_R(x)$ is a so-called sterile neutrino field, which does not carry any Standard Model quantum numbers.

The Lagrangian \eqref{Lagr_osc} is usually regarded as a classical object, with the fields $\nu_L$ and $\nu_R$ satisfying coupled equations of motion. The straightforward way to quantize the model starts by diagonalizing the Lagrangian, which becomes:
\begin{eqnarray}\label{Lagr_diag}
{\cal L}&=&\frac{1}{2}\overline{\nu}_1(x)(i\gamma^{\mu}\partial_{\mu}-m_1)\nu_1(x)+\frac{1}{2}\overline{\nu}_2(x)(i\gamma^{\mu}\partial_{\mu}-m_2)\nu_2(x),
\end{eqnarray}
upon the change of variables
\begin{equation}\label{cov}
\left(\begin{array}{c}
            \nu_L(x)\\
            C\bar\nu_R^T(x)
            \end{array}\right)=U\left(\begin{array}{c}
            \nu_{1L}(x)\\
            \nu_{2L}(x)
            \end{array}\right).
\end{equation}
The unitary matrix $U$ above has the form (Autonne-Takagi factorization)
\begin{equation}\label{mixing}
U=\left(\begin{array}{cc}
            \cos\theta&\sin\theta\\
           -\sin\theta&\cos\theta
            \end{array}\right)\left(\begin{array}{cc}
            \rho_1&0\\
          0&\rho_2
            \end{array}\right)
\end{equation}
where
\begin{equation}\label{theta}
\tan2\theta=\frac{2m_D}{m_R-m_L}
\end{equation}
and
\begin{equation}\label{masses}
m_{1,2}=\left|\frac{1}{2}(m_R-m_L)\mp\frac{1}{2}\sqrt{(m_R-m_L)^2+4m_D^2}\right|.
\end{equation}
The parameters $\rho_1$ and $\rho_2$ are chosen such that $\rho_i^2=\pm1$, with the role of compensating for a possible negative mass which can appear from the interplay of the values of the parameters $m_D,m_L,m_R$. For example, $\rho_2=1$ always, since $\frac{1}{2}(m_R-m_L)+\frac{1}{2}\sqrt{(m_R-m_L)^2+4m_D^2}>0$. On the other hand, $\rho_1=1$
if $m_Rm_L\geq m_D^2$, while $\rho_1=i$ if $m_Rm_L\leq m_D^2$.

The ordering $m_L\ll m_D\ll m_R$ insures the seesaw property: 
\begin{eqnarray}\label{ssmass}
m_1\approx\left|m_L-\frac{m_D^2}{m_R}\right|,\quad\quad m_2\approx m_R.
\end{eqnarray}
Typically, the left-handed Majorana mass $m_L$ is generated also by a seesaw effect, for example by coupling the Standard Model lepton doublets with a $SU(2)$ triplet of massive scalars, in which case $m_L\approx\lambda\frac{v^{2}}{\Lambda}$, where $\lambda$ is a coupling constant, $v$ is the vacuum expectation value of the Higgs field of the Standard Model and $\Lambda$ is a high-energy scale proportional to the mass of the triplet of scalars. When $m_L\equiv 0$, we have the famous seesaw mechanism, nowadays called a type I seesaw. When $m_D=m_R\equiv0$, we may have a type II seesaw. Irrespective of the details, a high energy scale beyond the Standard Model (typically a GUT scale of $10^{15}-10^{16}$ GeV) is considered to be responsible for the lightness of the observed neutrino masses in the seesaw mechanism\footnote{Due to the sensitivity of the quantum corrections of the Higgs particle mass to the high-energy seesaw scale, there appears inevitably a hierarchy problem, which implies a rather drastic fine tuning of the Yukawa couplings. On the other hand, there are models in which the high-energy scale is pushed down to TeV level or even lower, but in these cases the smallness of neutrino masses cannot be attributed at all to a seesaw effect.}. The type I and III seesaw imply the existence of sterile neutrinos and their mixing with the active species, while in type II seesaw sterile neutrinos do not appear. The question about the realization in nature of the seesaw mechanism in any of its versions is open. nevertheless there are stringent cosmological bounds on the mixture of active-sterile neutrinos, mainly from analyses of Big Bang Nucleosynthesis \cite{Dolgov-Barbieri, Dolgov-BBN, Dolgov-Villante} (see also \cite{Kainulainen}).

The fields $\nu_1$ and $\nu_2$ in \eqref{Lagr_diag} satisfy free Dirac equations with definite masses $m_1$ and $m_2$, as well as the constraints
\begin{eqnarray}\label{Majorana_cond}
\nu_i=C\bar\nu_i^T,\quad\quad i=1,2,
\end{eqnarray}
since, from \eqref{cov}, it follows that
\begin{eqnarray}
\nu_i=\nu_{iL}+C\bar\nu_{iL}^T,\quad\quad i=1,2.
\end{eqnarray}
The system described by the Lagrangian \eqref{Lagr_diag} is straightforwardly quantized canonically as two independent, free, Majorana fields of different masses (see formulas \eqref{massive_modes} below).

For three generation mixing, we may assume that there are three sterile\footnote{There is no stringent reason to assume that the number of sterile fields is equal to the number of active fields. We do it here in order to make the technical aspects more transparent, but all the results can be derived also with an arbitrary number of sterile fields.} right-handed neutrino fields $\nu_{s_iR},\ s_i=s_1,s_2,s_3$ apart from the three active flavour ones $\nu_{lL}, \ l=e,\mu,\tau$. The Lagrangian is
\begin{eqnarray}\label{Lagr_3}
{\cal L}&=&\overline{\nu}_L(x)i\gamma^{\mu}\partial_{\mu}\nu_L(x)+\overline{\nu}_R(x)i\gamma^{\mu}\partial_{\mu}\nu_R(x)\cr
& -& \overline{\nu}_L(x)M_D\nu_R(x)-\frac{1}{2}\bar\nu_L(x)\,M_L(\nu_L)^c(x) -\frac{1}{2}\bar\nu_R(x)\,M_R(\nu_R)^c(x) +h.c.,
\end{eqnarray}
where
\begin{eqnarray}\label{fields_3}
\nu_L=\left(\begin{array}{c}
            \nu_{eL}\\
           \nu_{\mu L}\\
\nu_{\tau L}
            \end{array}\right),\quad
\nu_R=\left(\begin{array}{c}
            \nu_{s_1R}\\
           \nu_{s_2R}\\
\nu_{s_3R}
            \end{array}\right),\quad
(\nu_L)^c=\left(\begin{array}{c}
            C\bar\nu_{eL}^T\\
           C\bar\nu_{\mu L}^T\\
C\bar\nu_{\tau L}^T
            \end{array}\right),\quad
(\nu_R)^c=\left(\begin{array}{c}
            C\bar\nu_{s_1R}^T\\
          C\bar \nu_{s_2R}^T\\
C\bar\nu_{s_3R}^T
            \end{array}\right),
\end{eqnarray}
and $M_D$, $M_L$, $M_R$ are $3\times 3$ complex non-diagonal mass matrices, the last two being symmetrical\footnote{The notation $(\nu_L)^c$ is just a short-hand and we do not attribute it the meaning of charge conjugation operation for the chiral fields, for reasons elaborated upon in \cite{pseudoC}.}. If we present the Lagrangian as
\begin{eqnarray}\label{Lagr_3'}
{\cal L}=\bar{n}_L(x)i\gamma^{\mu}\partial_{\mu}n_L(x)+\overline{({n}_L)^c}(x)i\gamma^{\mu}\partial_{\mu}{({n}_L)^c} -\frac{1}{2}\bar n_L(x)\,M (n_L)^c(x) -\frac{1}{2}\overline { (n_L)^c}(x)\,M n_L(x),
\end{eqnarray}
where
\begin{eqnarray}
n_L=\left(\begin{array}{c}
            \nu_L\\
            (\nu_R)^c
            \end{array}\right),\quad\quad M=\left(\begin{array}{cc}
            M_L&M_D\\
            M_D^T&M_R
            \end{array}\right),
\end{eqnarray}
the diagonalization is achieved by a unitary transformation
\begin{eqnarray}\label{rot_diag_3}
\nu_i(x)=V^\dagger n_L(x) = \left(\begin{array}{c}
            \nu_1 (x)\\
            \nu_2 (x)\\
\vdots\\
\nu_6 (x)
            \end{array}\right),
\end{eqnarray}
where $V$ is a $6\times 6$ unitary matrix, such that
\begin{equation}\label{mass_diag}
M=V m V^T, \ \ \mbox{with}\ \ m_{ik}=m_i\delta_{ik},\ \ m_i>0, \ \  \forall i=1,2,\ldots,6.
\end{equation}
The actual symbolic diagonalization of the $6\times 6$ complex symmetric matrix $M$ is very difficult, therefore in practical situations one makes simplifying assumptions. For example, in the case of type I seesaw, when $M_L\equiv 0$ and all the entries of the matrix $M_R$ are much larger in absolute value than all the entries of $M_D$, then one can bring the mass matrix $M$ to an approximate block diagonal form, with a $3\times 3$ matrix $M_{\small\mbox{light}}\approx-M_D^TM_R^{-1}M_D$ containing small mass parameters and a second $3\times 3$ matrix $M_{\small\mbox{heavy}}\approx M_R$ containing heavy masses. The bottom line is that three of the eigenvalues of the mass matrix $M$ are tiny and three are very large.

The fields $\nu_i,\ i=1,2,\ldots,6$, satisfy the Majorana condition $\nu_i=C\bar\nu_i^T$ and in terms of them the Lagrangian \eqref{Lagr_3'} is diagonal:
\begin{eqnarray}\label{Lagr_3diag}
{\cal L}&=&\frac{1}{2}\sum_{i=1}^6\overline{\nu}_i(x)(i\gamma^{\mu}\partial_{\mu}-m_i)\nu_i(x).
\end{eqnarray}
Again, the quantization of this system of free fields is trivial.

This procedure does not define unambiguously the flavour states associated to the flavour fields $\nu_L$ and $\nu_R$ which appear in \eqref{Lagr_3'}. The fields $\nu_L$ and $\nu_R$ are regarded as interacting ones, with bilinear interactions given by the mass terms in \eqref{Lagr_3'}. For interacting fields, one can not find a Fock representation \cite{SW, Strocchi}, consequently no "flavour states" associated to them \cite{AT_comment}. Traditionally, here comes the Gribov--Pontecorvo conjecture that such states would be produced coherently, in the form of \eqref{state_mix_N}, and subsequently oscillate. It is by now common knowledge that this is a phenomenological definition of flavour states, without an exact derivation from QFT principles. It rather mimics the two- or three-level systems in quantum mechanics; nevertheless, the principles of quantum mechanical coherent superpositions of states can not be applied to the superposition of {\it states of different systems}, i.e. particles of different mass \cite{AT_ICHEP}. In what follows, we shall present a quantum field theoretical construction of coherent flavour states for seesaw neutrinos.

\section{Oscillating neutrino states in the seesaw mechanism}\label{osc_seesaw}

The starting point of our proposed approach for defining coherent oscillating states is the general principle that all states of free particles in quantum field theory have to be generated by the action of a creation operator on the physical vacuum of the theory. In our prescription, the operator which fulfills this task is the creation operator of {\it massless} Standard Model neutrinos, acting on the vacuum of the {\it massive physical neutrinos}. 

%The analysis of particle states in quantum field theory can be done only in Hamiltonian formulation. 
We saw in Sec. \ref{Lagr_formalism} that the mixing of fields for diagonalizing the Lagrangian does not provide a definition of coherent mixtures of states. In this section, we shall show that the diagonalization of the Hamiltonian will give us a handle to define the flavour neutrino states. We shall explain the conceptual and technical details using the one-generation seesaw model \eqref{Lagr_osc} and generalize the results for the three-generation model.

It is known that the flavour fields included in the charged current weak interactions do not admit a Fock representation in theories with nondiagonal neutrino mass terms \cite{Bilenky-Giunti-flavour} (see also \cite{AT_comment}). According to Pontecorvo's conjecture that is the basis of the traditional approach to neutrino oscillations, the flavour states are postulated by implementing the same unitary transformation among the massive neutrino states. This definition would be rigorous {\it only if} the massive states have the same mass, but in this limit the oscillations vanish. The problem is further complicated by the requirement of coherence.

In order to define states by the action of an operator on the vacuum, we have to employ the Hamiltonian formulation. The results of Hamiltonian diagonalization have to agree with those of Lagrangian diagonalization. The difference is that in the Lagrangian formalism we first diagonalize and then quantize, while in the Hamiltonian formalism we start with the quantization and then diagonalize. 

\subsection{Two inequivalent Fock representations}

The classical Hamiltonian corresponding to the quadratic Lagrangian \eqref{Lagr_osc} is:
\begin{eqnarray}\label{Hamilt_osc}
{H}&=&\int d^3x\Big[-\overline{\nu}_L(x)i\gamma^{k}\partial_{k}\nu_L(x) -
\overline{\nu}_R(x)i\gamma^{k}\partial_{k}\nu_R(x)\cr
&+&m_D\big(\overline{\nu}_L(x)\nu_R(x)+\overline{\nu}_R(x)\nu_L(x)\big)\cr
&+&\frac{m_L}{2}\big(\nu_L^{T}(x)C\nu_L(x) +\overline{\nu}_L(x)C\overline{\nu}_L^{T}(x)\big)+\frac{m_R}{2}\big(\nu_R^{T}(x)C\nu_R(x) +\overline{\nu}_R(x)C\overline{\nu}_R^{T}(x)\big)\Big]\cr
&=&H_0+H_{mass}.
\end{eqnarray}
The fields $\nu_L(x)$ and $\nu_R(x)$ are coupled by their equations of motion,
therefore they have to be treated as {\it interacting fields}, with the interaction terms proportional to $m_D, m_L, m_R$: 
\begin{eqnarray}\label{coupled_eom}
&&i\gamma^{\mu}\partial_{\mu}\nu_{L}(x)=m_D\nu_R(x)+m_L \nu_L(x),\nonumber\\
&&i\gamma^{\mu}\partial_{\mu}\nu_{R}(x)=m_D\nu_{L}(x)+m_R\nu_R(x).
\end{eqnarray}
Let us recall that in the Standard Model, all fermionic fields are initially introduced as massless fields, and some of them acquire mass by the Yukawa interaction through the Brout--Englert--Higgs (BEH) mechanism. This argument is extended to any model based on gauge theories (GUT, for example), in which fermionic fields appear at the beginning as massless multiplet representation of some Lie groups, and acquire mass by spontaneous symmetry breaking (SSB). In the same vein, we consider the fields $\nu_L(x)$ and $\nu_R(x)$ as originally massless, all the mass terms $m_D, m_L, m_R$ appearing as effective expressions of an attractive interaction and SSB. 

On the other hand, the Hamiltonian corresponding to the Lagrangian \eqref{Lagr_diag} is
\begin{eqnarray}\label{H_diag}
{H}=\int d^3x&\big[-\overline{\nu}_1(x)i\gamma^{k}\partial_{k}\nu_1(x) +m_1\overline{\nu}_1(x)\nu_1(x)\cr
&-
\overline{\nu}_2(x)i\gamma^{k}\partial_{k}\nu_2(x)+m_2\overline{\nu}_2(x)\nu_2(x)\big],
\end{eqnarray}
where the two masses are given by \eqref{masses}, with the fields satisfying the free Dirac equations
\begin{eqnarray}\label{Dirac_eq}
&&(i\gamma^{\mu}\partial_{\mu}-m_1)\nu_{1}(x)=0,\nonumber\\
&&(i\gamma^{\mu}\partial_{\mu}-m_2)\nu_{2}(x)=0,
\end{eqnarray}
as well as the Majorana conditions \eqref{Majorana_cond}.

Since the Lagrangians \eqref{Lagr_osc} and \eqref{Lagr_diag} are equivalent, the Hamiltonians \eqref{Hamilt_osc} and \eqref{H_diag} must describe one and the same system. In the following, we shall exploit this equivalence.

The Hamiltonian \eqref{H_diag}, being in diagonal form, indicates that the system admits a Fock representation. The fields $\nu_1$ and $\nu_2$ are of Majorana type, and they are straightforwardly quantized canonically by imposing the equal-time anticommutation relations, which leads to the quantum fields
\begin{eqnarray}\label{massive_modes}
\nu_1(x)&=&\int\frac{d^3p}{(2\pi)^{3/2}\sqrt{2 E_{1\tp}}}\sum_\lambda\left(A_{1\lambda}({\bf p})U_\lambda(m_1,{\bf p})e^{-ipx}+A^\dagger_{1\lambda}({\bf p})V_\lambda(m_1,{\bf p})e^{ipx}\right),\cr
\nu_2(x)&=&\int\frac{d^3p}{(2\pi)^{3/2}\sqrt{2 E_{2\tp}}}\sum_\lambda\left(A_{2\lambda}({\bf p})U_\lambda(m_2,{\bf p})e^{-ipx}+A^\dagger_{2\lambda}({\bf p})V_\lambda(m_2,{\bf p})e^{ipx}\right),
\end{eqnarray}
where $ E_{i\tp}=\sqrt{\tp^2+m_i^2},\ i=1,2$ and $\lambda$ stands for the helicity. The spinors $U_\lambda(m_i,{\bf p}), V_\lambda(m_i,{\bf p})$ are defined in Appendix \ref{appendix1}.  The creation and annihilation operators satisfy the algebra
\begin{eqnarray}\label{ACR_A}
\{A_{i\lambda}({\bf p}),A^\dagger_{j\lambda'}({\bf k})\}=\delta_{ij}\delta_{\lambda\lambda'}\delta({\bf p}-{\bf k}),\end{eqnarray}
all the other anticommutators being zero. The Fock space of the model contains Majorana particle states with masses $m_1$ and $m_2$:
\begin{eqnarray}
A^\dagger_{1\lambda}({\bf p})|\Phi_0\rangle,\quad\quad A^\dagger_{2\lambda}({\bf p})|\Phi_0\rangle, \quad\quad\rm{etc.,}
\end{eqnarray}
where $|\Phi_0\rangle$ is the physical vacuum of the theory, satisfying
\begin{eqnarray}\label{vac_cond'}
A_{i\lambda}({\bf p})|\Phi_0\rangle=0,\ \ \ \ i=1,2.
\end{eqnarray}
The mode expansion of the normally-ordered Hamiltonian reads:
\begin{eqnarray}\label{H_osc_diag}
H&=\int d^3 p \sum_{\lambda} \Big[ E_{1\tp} A^\dagger_{1\lambda}({\bf p}) A_{1\lambda}({\bf p})+ E_{2\tp} A^\dagger_{2\lambda}({\bf p}) A_{2\lambda}({\bf p})\Big],
\end{eqnarray}
which satisfies the axiom
\begin{equation}\label{vac_cond}
H|\Phi_0\rangle=0.
\end{equation}

On the other hand, the Hamiltonian \eqref{Hamilt_osc} cannot be expanded in terms of creation and annihilation operators of the fields $\nu_L(x)$ and $\nu_R(x)$, because as interacting fields they do not admit a Fock representation (the separation into positive and negative energy modes is not relativistically invariant, since the fields do not satisfy the wave equation; see, e.g., \cite{Bog-Shirk}). What we {\it can} do, knowing that the Hamiltonian is constant in time, is to go to $t=0$, namely to consider $\nu_L({\bf x},0)$ and $\nu_R({\bf x},0)$. The time dependency can be restored by the Heisenberg equations of motion
\begin{equation}\label{H-S-connection}
\nu_{L,R}({\bf x},t)= e^{iHt }\nu_{L,R}({\bf x},0)e^{-iHt }.
\end{equation}
This is the standard procedure in the case of interacting fields, described, for example by Bjorken and Drell (see \cite{bjorken}, Sect. 15.4)\footnote{Regarding $\nu_{L,R}({\bf x},t)$ as Heisenberg fields and $\nu_{L,R}({\bf x},0)$ as Schr\"odinger fields, eq. \eqref{H-S-connection} represents the connection between the Heisenberg and Schr\"odinger pictures.}. Since $\nu_{L,R}({\bf x},t)$ as quantum fields have to satisfy the equal-time anticommutation relations, the procedure outlined above is equivalent to identifying at $t=0$ the interacting fields $\nu_{L,R}({\bf x},0)$ with the free massless fields $\psi_{L,R}({\bf x},0)$, which satisfy the Weyl equations:
\begin{equation}\label{weyl}
i\gamma^{\mu}\partial_{\mu}\psi_{L,R}(x)=0.
\end{equation}
Namely, we are allowed to write formally (see also \cite{NJL,UTK}):
\begin{eqnarray}\label{NJL}
\nu_L({\bf x},0)=\psi_L({\bf x},0),\nonumber\\
\nu_R({\bf x},0)=\psi_R({\bf x},0),
\end{eqnarray}
where, in the helicity basis (see Appendix \ref{appendix1} for the definitions of the spinors),
\begin{eqnarray}\label{Dirac_mode_exp}
\psi_L({\bf x},0)=\int\frac{d^3p}{(2\pi)^{3/2}\sqrt{2\tp}}\left(a_\da({\bf p})u_\da({\bf p})e^{i{\bf p\cdot x }}+b^\dagger_\ua({\bf p})v_\ua({\bf p})e^{-i{\bf p\cdot x }}\right),\nonumber\\
\psi_R({\bf x},0)=\int\frac{d^3p}{(2\pi)^{3/2}\sqrt{2\tp}}\left(c_\ua({\bf p})u_\ua({\bf p})e^{i{\bf p\cdot x }}+d^\dagger_\da({\bf p})v_\da({\bf p})e^{-i{\bf p\cdot x }}\right).
\end{eqnarray}
The operators $a_\da({\bf p}), b_\ua({\bf p}), c_\ua({\bf p}), d_\da({\bf p})$ annihilate the vacuum $|0\rangle$ of the fields  $\psi_{L,R}$, which we call the {\it massless neutrino vacuum},
\begin{eqnarray}\label{naive_vac}
a_\da({\bf p})|0\rangle=b_\ua({\bf p})|0\rangle=c_\ua({\bf p})|0\rangle=d_\da({\bf p})|0\rangle=0,\end{eqnarray}
and satisfy ordinary anticommutation relations:
\begin{eqnarray}\label{ACR_ord}
\{a_\da({\bf p}),a^\dagger_{\da}({\bf k})\}&=&\delta({\bf p}-{\bf k}),\quad\quad\quad\{c_\ua({\bf p}),c^\dagger_{\ua}({\bf k})\}=\delta({\bf p}-{\bf k}),\nonumber\\
\{b_\ua({\bf p}),b^\dagger_{\ua}({\bf k})\}&=&\delta({\bf p}-{\bf k}),\quad\quad\quad
\{d_\da({\bf p}),d^\dagger_{\da}({\bf k})\}=\delta({\bf p}-{\bf k}),
\end{eqnarray}
all the other anticommutators being zero. The states 
\begin{eqnarray}\label{bare_neutrinoLL}
a^\dagger_\da({\bf p})|0\rangle \ \ \ \text{and }\ \ \ \ b^\dagger_\ua({\bf p})|0\rangle\end{eqnarray}
represent the massless left-helicity neutrino and right-helicity antineutrino states, corresponding to the active field $\psi_L$. Since the field $\psi_L(x)$ coincides with the Standard Model flavour neutrino field, to the states \eqref{bare_neutrinoLL} we assign the lepton numbers $+1$ and $-1$, respectively. The states
\begin{eqnarray}\label{bare_neutrino_R}
c^\dagger_\ua({\bf p})|0\rangle \ \ \ \text{and }\ \ \ \ d^\dagger_\da({\bf p})|0\rangle\end{eqnarray}
represent the massless right-helicity neutrino and left-helicity antineutrino, corresponding to the bare sterile field $\psi_R$. Altogether, the states created from the vacuum $|0\rangle$ form the Fock representation of the massless Standard Model neutrino fields.

We emphasize once more that the operators $a_\da({\bf p}), b_\ua({\bf p}), c_\ua({\bf p}), d^\dagger_\da({\bf p})$ are  not annihilation operators for the physical vacuum $|\Phi_0\rangle$ defined by \eqref{vac_cond}.

Using \eqref{Dirac_mode_exp}  in \eqref{Hamilt_osc}, we find, with the help of the relations \eqref{spinor_rel}:
\begin{eqnarray}\label{H_modes}
H&=&\int d^3 p\ \tp\left(a^\dagger_\da({\bf p}) a_\da({\bf p})+b^\dagger_\ua({\bf p})b_\ua({\bf p})+c^\dagger_\ua({\bf p})c_\ua({\bf p})+d^\dagger_\da({\bf p}) d_\da({\bf p})\right)\cr
&+&i\int d^3 p \Big[m_D\left(a^\dagger_\da({\bf p}) d^\dagger_\da(-{\bf p})+d_\da({\bf p}) a_\da(-{\bf p})-c^\dagger_\ua({\bf p}) b^\dagger_\ua(-{\bf p})-b_\ua({\bf p}) c_\ua(-{\bf p})\right)\cr
&-&\frac{m_L}{2}\left( a^\dagger_\da({\bf p}) a^\dagger_\da(-{\bf p})+a_\da({\bf p}) a_\da(-{\bf p})+b^\dagger_\ua({\bf p}) b^\dagger_\ua(-{\bf p})+b_\ua({\bf p}) b_\ua(-{\bf p}) \right)\cr
&-&\frac{m_R}{2}\left( c^\dagger_\ua({\bf p}) c^\dagger_\ua(-{\bf p})+c_\ua({\bf p}) c_\ua(-{\bf p})+d^\dagger_\da({\bf p}) d^\dagger_\da(-{\bf p})+d_\da({\bf p}) d_\da(-{\bf p}) \right) \Big].
\end{eqnarray}
One can verify that
\begin{equation}
H_0|0\rangle= 0,
\end{equation}
but 
\begin{equation}
H|0\rangle\neq 0,
\end{equation}
which confirms once more that the bare vacuum $|0\rangle$ and the physical vacuum $|\Phi_0\rangle$ are different.

To summarize, we have expressed the Hamiltonian of see-saw mechanism, corresponding to the Lagrangian \eqref{Lagr_osc}, in two equivalent forms, \eqref{H_osc_diag} and \eqref{H_modes}. We have used for this two {\it inequivalent} sets of creation and annihilation operators, corresponding  i) to the Fock representations of massive neutrinos with the vacuum $|\Phi_0\rangle$ and ii) to the massless SM neutrinos, with the vacuum $|0\rangle$. Between the two sets of operators there are relations, called Bogoliubov transformations, that we will establish using \eqref{cov}, \eqref{massive_modes}, \eqref{NJL} and \eqref{Dirac_mode_exp}.

\subsection{Bogoliubov transformations}

To find the Bogoliubov transformations\footnote{This is analogous to the procedure used in the Nambu--Jona-Lasinio model according to the original paper \cite{NJL}. The essential difference is that in the case of neutrinos, the massless fields will need to be mixed before applying the scheme.}, we use first \eqref{NJL} to define a new set of massless free fields, $\psi_{1L}(x),\psi_{2L}(x)$, by the following sequence of equalities:
\begin{equation}\label{rot1}
\left(\begin{array}{c}
            \nu_L({\bf x},0)\\
            C\bar\nu_R^T({\bf x},0)
            \end{array}\right)=\left(\begin{array}{c}
            \psi_L({\bf x},0)\\
            C\bar\psi_R^T({\bf x},0)
            \end{array}\right)=U\left(\begin{array}{c}
            \psi_{1L}({\bf x},0)\\
            \psi_{2L}({\bf x},0)
            \end{array}\right),
\end{equation}
with the mixing matrix $U$ given by \eqref{mixing}. As massless chiral fields, $\psi_{1L}(x),\psi_{2L}(x)$
satisfy also the Weyl equation
\begin{equation}\label{Weyl_psi}
i\gamma^\mu\partial_\mu\psi_{iL}(x)=0,\ \ \ i=1,2
\end{equation}
and are expanded as:
\begin{eqnarray}\label{psi_modes}
\psi_{1L}(x)
&=&\int\frac{d^3p}{(2\pi)^{3/2}\sqrt{2\tp}}\left(a_{1\da}({\bf p})u_\da({\bf p})e^{-ipx}+a^\dagger_{1\ua}({\bf p})v_\ua({\bf p})e^{ipx}\right),\cr
\psi_{2L}(x)
&=&\int\frac{d^3p}{(2\pi)^{3/2}\sqrt{2\tp}}\left(a_{2\da}({\bf p})u_\da({\bf p})e^{-ipx}+a^\dagger_{2\ua}({\bf p})v_\ua({\bf p})e^{ipx}\right).
\end{eqnarray}
Using the second equality in \eqref{rot1} and \eqref{psi_modes}, we find immediately:
\begin{equation}\label{rot_ad}
\left(\begin{array}{c}
            a_{1\da}({\bf p})\\
            a_{2\da}({\bf p})
            \end{array}\right)=U^{-1}\left(\begin{array}{c}
            a_{\da}({\bf p})\\
            -d_\da({\bf p})
            \end{array}\right),
\end{equation}
\begin{equation}\label{rot_bc}
\left(\begin{array}{c}
            a_{1\ua}^\dagger({\bf p})\\
           a_{2\ua}^\dagger({\bf p})
            \end{array}\right)=U^{-1}\left(\begin{array}{c}
            b_{\ua}^\dagger({\bf p})\\
            c_\ua^\dagger({\bf p})
            \end{array}\right).
\end{equation}
It is straightforward to check, using \eqref{naive_vac} and \eqref{ACR_ord}, that
\begin{eqnarray}\label{ACR_psi}
\{a_{1\lambda}({\bf p}),a^\dagger_{1\lambda'}({\bf k})\}&=&\delta_{\lambda\lambda'}\delta({\bf p}-{\bf k}),\nonumber\\
\{a_{2\lambda}({\bf p}),a^\dagger_{2\lambda'}({\bf k})\}&=&\delta_{\lambda\lambda'}\delta({\bf p}-{\bf k}),
\end{eqnarray}
all the other anticommutators being zero, as well as
\begin{eqnarray}
a_{1\lambda}({\bf p})|0\rangle=a_{2\lambda}({\bf p})|0\rangle=0.
\end{eqnarray}
The rotations  \eqref{rot_ad} and \eqref{rot_bc} are transformations in the Fock space of massless particles, mixing massless neutrinos (active and sterile) with definite lepton number into massless neutrinos with undefined lepton number. This is a genuine change of basis in the one-particle Hilbert space that underlies the Fock space.
We emphasize that by such a transformation we obtain a coherent superposition of particle states of {\it identical mass}\footnote{This argument is used by Gell-Mann and Pais in their seminal paper on CP violation in the $K_0-\bar K_0$ system \cite{GM-P}, and reiterated by Pais and Piccioni \cite{Pais-Piccioni} in the pioneering paper on the particle-antiparticle oscillations of neutral kaons.}.

Then we rewrite \eqref{cov} at $t=0$:
\begin{equation}\label{rot2}
\left(\begin{array}{c}
            \nu_L({\bf x},0)\\
            C\bar\nu_R^T({\bf x},0)
            \end{array}\right)=U\left(\begin{array}{c}
            \nu_{1L}({\bf x},0)\\
            \nu_{2L}({\bf x},0)
            \end{array}\right).
\end{equation}
Comparing \eqref{rot1} with \eqref{rot2}, we find
\begin{equation}\label{NJL'}
\left(\begin{array}{c}
             \psi_{1L}({\bf x},0)\\
            \psi_{2L}({\bf x},0)
            \end{array}\right)=\left(\begin{array}{c}
            \nu_{1L}({\bf x},0)\\
            \nu_{2L}({\bf x},0)
            \end{array}\right),
\end{equation}
where $\psi_{iL}$ are free massless fields and $\nu_{iL}$ are L-chiral components of free massive Majorana fields. Naturally, eq. \eqref{NJL'} is satisfied operatorially {\it only} at $t=0$, because the two sets of fields evolve in time with different Hamiltonians. 

Using the mode-expansions \eqref{massive_modes} and \eqref{psi_modes} in \eqref{NJL'}, we find the equations
\begin{eqnarray}
\sum_\lambda\left(A_{1\lambda}({\bf p})U_\lambda(m_1,{\bf p})+A^\dagger_{1\lambda}(-{\bf p})V_\lambda(m_1,-{\bf p})\right)={\sqrt\frac{ E_{1\tp}}{\tp}}\sum_\lambda\left(a_{1\lambda}({\bf p})u_\lambda({\bf p})+\sgn\lambda a^\dagger_{1\lambda}(-{\bf p})v_\lambda(-{\bf p})\right),\cr
\sum_\lambda\left(A_{2\lambda}({\bf p})U_\lambda(m_2,{\bf p})+A^\dagger_{2\lambda}(-{\bf p})V_\lambda(m_2,-{\bf p})\right)={\sqrt\frac{ E_{2\tp}}{\tp}}\sum_\lambda\left(a_{2\lambda}({\bf p})u_\lambda({\bf p})+\sgn\lambda a^\dagger_{2\lambda}(-{\bf p})v_\lambda(-{\bf p})\right),\cr
\end{eqnarray}
where we used the Majorana-extensions $\psi_i=\psi_{iL}+C\bar\psi_{iL}^T$, $i=1,2$. Multiplying from the left the first equation by $U^\dagger_{\lambda'}(m_1,{\bf p})$ and the second by $U^\dagger_{\lambda'}(m_2,{\bf p})$, we obtain:
\begin{eqnarray}\label{BT}
A_{1\lambda}({\bf p})=\frac{1}{2\sqrt{ E_{1\tp}\tp}}U^\dagger_{\lambda}(m_1,{\bf p})u_\lambda({\bf p})a_{1\lambda}({\bf p})+\sgn\lambda\,U^\dagger_{\lambda}(m_1,{\bf p}) v_\lambda(-{\bf p})a^\dagger_{1\lambda}(-{\bf p}),\cr
A_{2\lambda}({\bf p})=\frac{1}{2\sqrt{ E_{2\tp}\tp}}U^\dagger_{\lambda}(m_2,{\bf p})u_\lambda({\bf p})a_{2\lambda}({\bf p})+\sgn\lambda\,U^\dagger_{\lambda}(m_2,{\bf p}) v_\lambda(-{\bf p})a^\dagger_{2\lambda}(-{\bf p}).
\end{eqnarray}
Using the formulas in Appendix \ref{appendix1}, we bring the Bogoliubov transformations to the form:
\begin{eqnarray}\label{Btrans}
A_{1\lambda}({\bf p})&=&\alpha_{1\tp}a_{1\lambda}({\bf p})+i\beta_{1\tp}\,a^\dagger_{1\lambda}(-{\bf p}),\cr
A_{2\lambda}({\bf p})&=&\alpha_{2\tp}a_{2\lambda}({\bf p})+i\beta_{2\tp}\,a^\dagger_{2\lambda}(-{\bf p}),
\end{eqnarray}
where $\alpha_{i\tp},\ \beta_{i\tp}$, $i=1,2$,  are real coefficients of the form
\begin{eqnarray}\label{BT_coeff}
\alpha_{i\tp}=\sqrt{\frac{ E_{i\tp}+\tp}{2 E_{i\tp}}},\ \ \ \ \ 
\beta_{i\tp}=\sqrt{\frac{ E_{i\tp}-\tp}{2 E_{i\tp}}},
\end{eqnarray}
satisfying 
\begin{eqnarray}\label{square_1_osc}
|\alpha_{i\tp}|^2+|\beta_{i\tp}|^2=1,\ \ \ i=1,2.
\end{eqnarray}
This ensures the compatibility between the canonical anticommutation relations \eqref{ACR_A} and \eqref{ACR_psi}.

Using the Bogoliubov transformations \eqref{Btrans}, we can find a formal relation between the bare vacuum $|0\rangle$ and the physical vacuum $|\Phi_0\rangle$.
The latter is a superposition of zero-momentum and zero-spin pairs of massless bare neutrinos:
\begin{eqnarray}
|\Phi_0\rangle ={\cal N}\ \Pi_{{\bf p},\lambda}\ e^{R_{1\tp}\,a_{1\lambda}^\dagger({\bf p})a_{1\lambda}^\dagger(-{\bf p})}e^{R_{2\tp}\,a_{2\lambda}^\dagger({\bf p})a_{2\lambda}^\dagger(-{\bf p})}|0\rangle.
\end{eqnarray}
The normalization constant $\cal N$ and the coefficients $R_{1\tp}$ and $R_{2\tp}$ are determined by using \eqref{vac_cond'} and the normalization condition $\langle\Phi_0|\Phi_0\rangle=1$. Taking into account also the fermionic nature of the particles (the creation operators are nilpotent), we obtain:
\begin{eqnarray}\label{normalized vacuum}
|\Phi_0\rangle =\ \Pi_{{\bf p},\lambda}\  \left(\alpha_{1\tp} -i\beta_{1\tp}\,a_{1\lambda}^\dagger({\bf p})a_{1\lambda}^\dagger(-{\bf p})\right)\left(\alpha_{2\tp} -i\beta_{2\tp}\,a_{2\lambda}^\dagger({\bf p})a_{2\lambda}^\dagger(-{\bf p})\right)|0\rangle.
\end{eqnarray}
If we express the products $a_{1\lambda}^\dagger({\bf p})a_{1\lambda}^\dagger(-{\bf p})$ and $a_{2\lambda}^\dagger({\bf p})a_{2\lambda}^\dagger(-{\bf p})$ in terms of the original operators $a^\dagger_\da({\bf p}), b^\dagger_\ua({\bf p}), c^\dagger_\ua({\bf p}),d^\dagger_\da({\bf p})$ using the inverse of \eqref{rot_ad} and \eqref{rot_bc}, we can see clearly that the vacuum state $|\Phi_0\rangle$ violates the lepton number symmetry. At the same time, $|\Phi_0\rangle$ preserves translational symmetry (the products have total momentum zero) and rotational invariance (the products have total spin zero). The Hamiltonian $H$ in the form \eqref{H_osc_diag} annihilates it. The state $|\Phi_0\rangle$ therefore satisfies the conditions for a physical vacuum \cite{SW}.

Although the fields $\psi_{1L}(x),\psi_{2L}(x)$ are massless, therefore Weyl, they can be regarded as "proto-Majorana" fields, in the sense that their creation and annihilation operators are directly connected by the Bogoliubov transformations \eqref{Btrans} to those of the Majorana fields $\nu_1(x),\nu_2(x)$.

The overlap of the two vacua, $|0\rangle$ and $|\Phi_0\rangle$, is
\begin{eqnarray}
\langle 0|\Phi_0\rangle = \Pi_{i,{\bf p},\lambda}\ \alpha_{i\tp}=\Pi_{i,{\bf p},\lambda}\ \left[\frac{1}{2}\left(1+\frac{\tp}{ E_{i\tp}}\right)\right]^{1/2}=\exp\left\{\sum_{i,{\bf p},\lambda}\frac{1}{2}\ln\left[\frac{1}{2}\left(1+\frac{\tp}{ E_{i\tp}}\right)\right]\right\},
\end{eqnarray}
which vanishes as $\exp\left[-V\pi(m_1^2+m_2^2)\int d{ \bf p}/(2\pi)^3\right]$ in the infinite volume and infinite momentum limit:
\begin{equation}
\langle 0|\Phi_0\rangle \to 0.
\end{equation}
In other words, the Fock space of massless flavour neutrinos is orthogonal to the Fock space of massive neutrinos. The two Fock representations of the canonical algebra are unitarily inequivalent. 

In Appendix \ref{diag}, we confirm the results by starting from the nondiagonal form of the Hamiltonian \eqref{H_modes} and bringing it to \eqref{H_osc_diag}, without using at any point the information on mixing matrices and mass parameters obtained from the Lagrangian diagonalization. Although the method given in Appendix \ref{diag} emphasizes the physical meaning of the procedure, it is more tedious. In the general case of three generations, it becomes quite hard to apply. On the other hand, the method presented in this subsection is straightforward to generalize, as is done in the next subsection.

For the one-generation type I seesaw case, the analysis stops here. There are no oscillating particle states, because flavour oscillations cannot take place with a single flavour. Oscillations into sterile states involve a huge mass difference between the Majorana fields, which presents an unobservable oscillation frequency. However, the technical aspects detailed above are essential for the formulation of the oscillating neutrino states in the next subsection.

\subsection{Three-generation seesaw model and oscillations}

In the case of three-generation mixing, we can define oscillating neutrino states. Just as in the case of Lagrangian diagonalization, the concrete analytical expressions of the $6 \times 6$ mixing matrix and of the diagonal mass matrix are practically impossible to find. Nevertheless, as it will be seen, the symbolical calculations lead to exact results, up to the precise form of the mass parameters. We consider the most general case, which allows also for CP violation.

We start with the Lagrangian \eqref{Lagr_3} and find the corresponding Hamiltonian. We treat the fields $\nu_L$ and $\nu_R$ in \eqref{fields_3} as chiral massless fields, with interactions given by the mass terms. Just as in the one-generation case (see eq. \eqref{NJL}), we identify
\begin{eqnarray}\label{NJL_3}
\nu_{lL}({\bf x},0)=\psi_{lL}({\bf x},0),\ \ \ \ l=e,\mu,\tau,\nonumber\\
\nu_{s_{i}R}({\bf x},0)=\psi_{s_{i}R}({\bf x},0),\ \ \ \ i=1,2,3.
\end{eqnarray}
Assuming the mode expansions of the bare active and sterile massless fields to be
\begin{eqnarray}\label{Dirac_mode_3}
\psi_{lL}({\bf x},0)=\int\frac{d^3p}{(2\pi)^{3/2}\sqrt{2\tp}}\left(a_{l\da}({\bf p})u_\da({\bf p})e^{i{\bf p\cdot x }}+b^\dagger_{l\ua}({\bf p})v_\ua({\bf p})e^{-i{\bf p\cdot x }}\right),\nonumber\\
\psi_{s_{i}R}({\bf x},0)=\int\frac{d^3p}{(2\pi)^{3/2}\sqrt{2\tp}}\left(c_{s_{i}\ua}({\bf p})u_\ua({\bf p})e^{i{\bf p\cdot x }}+d^\dagger_{s_{i}\da}({\bf p})v_\da({\bf p})e^{-i{\bf p\cdot x }}\right),
\end{eqnarray}
the Hamiltonian will have, analogously to \eqref{H_modes}, a non-diagonal form in the massless creation and annihilation operators. For its diagonalization, one defines first proto-Majorana massless fields analogous to \eqref{rot1}, by the rotation
\begin{equation}\label{rot1_3}
           \left(\begin{array}{c}
            \psi_{lL}({\bf x},0)\\
            C\bar\psi_{s_{i}R}^T({\bf x},0)
            \end{array}\right)=V\left(\begin{array}{c}
            \psi_{1L}({\bf x},0)\\
            \psi_{2L}({\bf x},0)\\
\vdots\\
\psi_{6L}({\bf x},0)
            \end{array}\right),
\end{equation}
where the matrix $V$ is the one that diagonalizes the Lagrangian (see eq. \eqref{rot_diag_3}), such that
\begin{eqnarray}\label{psi_modes_3}
\psi_{iL}(x)
&=&\int\frac{d^3p}{(2\pi)^{3/2}\sqrt{2\tp}}\left(a_{i\da}({\bf p})u_\da({\bf p})e^{-ipx}+a^\dagger_{i\ua}({\bf p})v_\ua({\bf p})e^{ipx}\right),\ i=1,2,\ldots,6.
\end{eqnarray}
From (\ref{Dirac_mode_3} -- \ref{psi_modes_3}),  we obtain the relations
\begin{eqnarray}\label{rot_abcd}
a_{i\da}({\bf p})&=&\sum_{l=e,\mu\tau}V^\dagger_{il}a_{l\da}({\bf p})+\sum_{j=1,2,3}V^\dagger_{is_{j}}\left(-d_{s_{j}\da}({\bf p})\right),\cr
a^\dagger_{i\ua}({\bf p})&=&\sum_{l=e,\mu\tau}V^\dagger_{il}b_{l\ua}^\dagger({\bf p})+\sum_{j=1,2,3}V^\dagger_{is_{j}}c_{s_{j}\ua}^\dagger({\bf p}),\ \ \ i=1,2,\ldots,6.
\end{eqnarray}
The next step is to make use of \eqref{rot_diag_3} and \eqref{rot1_3}, in order to relate the massless proto-Majorana fields $\psi_{i}(x)$ and the massive Majorana fields $\nu_{i}(x)$ at $t=0$:
\begin{equation}\label{NJL'_3}
             \psi_{i}({\bf x},0)=
            \nu_{i}({\bf x},0),\ \ \ \ i=1,2,\ldots,6,
\end{equation}
where 
\begin{equation}
\psi_{i}({\bf x},0)=\psi_{iL}({\bf x},0)+C\bar \psi_{iL}^T({\bf x},0)
\end{equation}
and
 \begin{eqnarray}\label{Majorana_mod_3}
\nu_i(x)=\int\frac{d^3p}{(2\pi)^{3/2}\sqrt{2 E_{i\tp}}}\sum_\lambda\left(A_{i\lambda}({\bf p})U_\lambda(m_i,{\bf p})e^{-ipx}+A^\dagger_{i\lambda}({\bf p})V_\lambda(m_i,{\bf p})e^{ipx}\right), 
\end{eqnarray}
with $ E_{i\tp}=\sqrt{\tp^2+m_i^2},$ and
\begin{eqnarray}
\{A_{i\lambda}({\bf p}),A^\dagger_{j\lambda'}({\bf k})\}=\delta_{ij}\delta_{\lambda\lambda'}\delta({\bf p}-{\bf k}),
\end{eqnarray}
all the other anticommutators being zero. The masses $m_i$ are the elements of the diagonal mass matrix defined in \eqref{mass_diag}.
Using \eqref{NJL'_3}, \eqref{Majorana_mod_3} and \eqref{psi_modes_3}, we find immediately the Bogoliubov transformations:
\begin{eqnarray}\label{Btrans_3}
A_{i\lambda}({\bf p})&=&\alpha_{i\tp}a_{i\lambda}({\bf p})+i\beta_{i\tp}\,a^\dagger_{i\lambda}(-{\bf p}), \ \ \ i=1,2,\ldots,6,
\end{eqnarray}
with 
\begin{eqnarray}\label{BT_coeff}
\alpha_{i\tp}=\sqrt{\frac{ E_{i\tp}+\tp}{2 E_{i\tp}}},\ \ \ \ \ 
\beta_{i\tp}=\sqrt{\frac{ E_{i\tp}-\tp}{2 E_{i\tp}}}.
\end{eqnarray}
The flavour number violating physical vacuum $|\Phi_0\rangle$, satisfying $A_{i\lambda}({\bf p})|\Phi_0\rangle=0, i=1,2,\ldots,6$, is formally a condensate of spinless and zero-momentum pairs of massless neutrinos and anti-neutrinos:
\begin{eqnarray}\label{normalized vacuum 3}
|\Phi_0\rangle =\ \Pi_{{\bf p},\lambda, i}\  \left(\alpha_{i\tp} -i\beta_{i\tp}\,a_{iM\lambda}^\dagger({\bf p})a_{iM\lambda}^\dagger(-{\bf p})\right)|0\rangle.
\end{eqnarray}

\subsection*{Definition of the oscillating neutrino states} 

Our proposed prescription for defining the oscillating neutrino states is to use the action of the SM flavour neutrino operators $a_{l\da}^\dagger({\bf p}), b_{l\ua}^\dagger({\bf p})$ on the physical vacuum $|\Phi_0\rangle$. Through the Schr\"odinger picture identification \eqref{NJL_3}, these operators are connected to the Majorana neutrino field $\nu_{lL}(x)$. The physical motivation for this choice is that those operators, acting on their own Fock space vacuum $|0\rangle$, create the SM neutrino states, which carry the flavour quantum number. Moreover, in the limit in which the lepton number-violating and flavour-violating interaction vanishes, the proposed oscillating neutrino states reduce to the SM flavour states, as the vacuum $\vert \Phi_0 \rangle$ also reduces to the vacuum $\vert 0 \rangle$.

We emphasize once more that the operators $a_{l\da}^\dagger({\bf p})$ cannot create one-particle Fock states from the physical vacuum. Instead they create coherent superpositions of massive Fock states. By inverting \eqref{rot_abcd} and \eqref{Btrans_3}, we find
\begin{eqnarray}\label{coh_state_mix_3}
|\nu_{l\da} ({\bf p})\rangle&=&a_{l\da}^\dagger({\bf p})|\Phi_0 \rangle=\sum_{i=1}^6 \alpha_{i\tp} V^*_{li} A^\dagger_{i\da}({\bf p})|\Phi_0\rangle=\sum_{i=1}^6\alpha_{i\tp} V^*_{li} |\nu_{i \da}({\bf p})\rangle,\cr
|\nu_{l\ua} ({\bf p})\rangle&=&b_{l\ua}^\dagger({\bf p})|\Phi_0 \rangle=\sum_{i=1}^6 \alpha_{i\tp}V_{li} A^\dagger_{i\ua}({\bf p})|\Phi_0\rangle=\sum_{i=1}^6 \alpha_{i\tp} V_{li}|\nu_{i \ua}({\bf p})\rangle.
\end{eqnarray}

If we introduce the diagonal matrix
\begin{equation}\label{alpha}
{\boldsymbol \alpha}(\tp)=
            \alpha_{i\tp}\delta_{ik}, \ \ \ \ i,k=1,2,\ldots,6,
\end{equation}
with $ \alpha_{i\tp}$ given by \eqref{BT_coeff}, then, if the massive fields are mixed with the unitary mixing matrix $V$, the massive states are mixed coherently with the non-unitary mixing matrix ${\boldsymbol \alpha}(\tp) V^*$ in order to create left-helicity neutrinos and with the matrix ${\boldsymbol \alpha}(\tp) V$ to create right-helicity (anti)neutrinos.
One can confirm that the oscillating neutrino/antineutrino states of different flavour and same momentum overlap, even without taking into account the smallness of the mixing coefficients of the heavy states:
\begin{equation}\label{nonorth}
\langle\nu_{l'\lambda} ({\bf p})|\nu_{l\lambda} ({\bf p})\rangle \neq 0,
\end{equation}
and their norm is subunitary:
\begin{equation}\label{nonnorm}
\langle\nu_{l\lambda} ({\bf p})|\nu_{l\lambda} ({\bf p})\rangle = \sum_i\alpha^2_{i\tp}|V_{li}|^2.
\end{equation}
In principle, the oscillating states can be normalized, but at this point we do not see any physical reason to do it. Note that the overlap of different states, though negligible with respect to the accuracy of the present oscillation experiments in the ultrarelativistic limit of the light neutrinos, cannot be removed via normalization. The overlap is a direct consequence of the coherence of the states.

The properties \eqref{nonorth} and \eqref{nonnorm} underline the fact that the relations \eqref{coh_state_mix_3} do not represent a change of basis between the mass eigenstates and some "weak eigenstates."

For type II seesaw, when there are only three light neutrinos and no super-massive states, the unitary mixing matrix $V$ is $3\times 3$ and the transition probability for flavour oscillations is given in the ultrarelativistic limit by
\begin{eqnarray}
{\cal P}_{\nu_l\to \nu_l'}^\textrm{II}=|\langle \nu_l({\bf p})|\nu_{l'}({\bf p})\rangle|^2 \approx \sum_{k,j=1}^3\alpha^2_{k\tp}\alpha^2_{j\tp}V^*_{lk}V_{l'k}V_{lj}V^*_{l'j} \exp{\left(-i\frac{\Delta m_{kj}^2 L}{2E}\right)},
\end{eqnarray}
where $E$ is the energy of the neutrinos and $L$ is the source-detector distance.

For type I seesaw, the contribution of the very heavy neutrino states to the transition probability is vanishingly small, because the coefficients $V_{lk}$, when $m_k$ is very large, are negligible compared to the mixing coefficients for the light states. The multiplication by $\alpha_{k\tp}$ reduces $V_{lk}$ even more (effectively by a factor $1/\sqrt 2$, when $\tp\ll m_k$).
Thus, we have
\begin{eqnarray}
{\cal P}_{\nu_l\to \nu_l'}^\textrm{I} \approx \sum_{k,j=1}^3\alpha^2_{k\tp}\alpha^2_{j\tp}N^*_{lk}N_{l'k}N_{lj}N^*_{l'j} \exp{\left(-i\frac{\Delta m_{kj}^2 L}{2E}\right)},
\end{eqnarray}
where the matrix $N$ is the $3\times 3$ upper left quadrant of the matrix $V$, namely the part which mixes the light neutrino fields. 

As far as CP violation is concerned, let us note that the real "coherence factors" $\alpha_{i\tp}$ do not modify the Dirac or Majorana CP violating phases, which are solely due to the fields mixing matrix $V$. Nevertheless, the absolute values of the mixing coefficients of states are modified depending on the momenta of the neutrinos. The analysis of CP violation with either Dirac or Dirac and Majorana phases remains the same as in the traditional approach \cite{B-H-P, Doi, Bernabeu-Pascual, deG-K-M, BPP, Xing_CP}.

\section{Discussion and outlook}\label{outlook}

The current neutrino oscillation paradigm asserts that neutrinos undergo flavour oscillations if the flavour states are superpositions of mass eigenstates and if the superpositions are coherent. In this paper, we have explored the prescription for defining intrinsically coherent oscillating neutrino states earlier proposed in \cite{AT_neutron, AT_neutrino}, within the context of the seesaw mechanism. The Majorana character of the particles that mix adds an extra layer of complication, as the vacuum violates fermionic number, as well as flavour. We have thus confirmed that the coherent flavour neutrino states have the universal and simple form \eqref{coh_state_mix_N}, irrespective of the Dirac or Majorana character of the neutrinos. The coherence factors are Bogoliubov coefficients relating the two natural Fock spaces (one fictitious -- the massless flavour space, and one physical -- the massive neutrino space) involved in the definition of the respective oscillating states.

In essence, the possibility of defining the oscillating neutrino states according to the present scheme is directly related to the structure of the vacuum of the physical massive neutrinos, as a condensate of Cooper-like pairs of massless SM flavour neutrinos. Expressed in this way, the vacuum clearly exhibits the fermion number violation, as well as the flavour violation, while retaining its Poincar\'e invariance, as it should. This is a manifestation of Coleman's theorem \cite{Coleman}, according to which in a relativistic quantum field theory, the invariance of the vacuum is the invariance of the world. The condensation of massless neutrinos is formally achieved by the attractive Yukawa interaction that leads to mass generation (either Dirac mass, through the SM Brout--Englert-Higgs mechanism, or Majorana mass, through a triplet Higgs interaction, as in type II seesaw). The interplay of two vacua and the interpretation of the Majorana neutrino as a Bogoliubov quasiparticle have been explored also in earlier papers \cite{Chang,FT2,FT3,KF_ss_vac}.

Unlike the quantum mechanical coherent states \cite{Klauder, Sudarshan,Glauber}, which are superpositions of an infinite number of particle states, the oscillating coherent states of neutrinos are not eigenstates of the annihilation operator. On the other hand, they share the characteristic non-orthogonality, meaning that two coherent states are never completely disjoint. As a result, there exists the so-called zero-distance flavour conversion, where any electron neutrino state, for example, has a tiny overlap with the muon neutrino and tau neutrino. Such an effect is well-known in the case of type I (and III) seesaw scheme (see, for example, \cite{Valle,Tortola}), but in our prescription it appears for all the Majorana or Dirac neutrino oscillations. A similar general overlap at zero-distance is encountered in the phenomenological definition of production and detection flavour neutrino states \cite{GKLL, pheno_states}. The amount of non-orthogonality of the coherent oscillating states in the present scheme depends on the energy of the states, being small for ultrarelativistic neutrino. Since hitherto the oscillation experiments have been performed with ultrarelativistic neutrinos, such departures from unitarity are far beyond the sensitivity of the current observations.

In this connection, we would like to emphasize that zero-distance conversion, even in the absence of sterile neutrinos, does not violate any physical principle. Since the neutrinos are not directly observed in experiments, it is customary to identify the flavour neutrinos by the emitted charged lepton, which {\it are} detected (see, for example, \cite{Bilenky-Giunti-flavour}). This is a justifiable phenomenological approach for interpreting the experimental data within the present accuracy levels. Nevertheless, it is a working definition and not a law. The fact that zero-distance conversion has not been observed is due to experimental limitations, and not to a physical principle which would {\it impose} that flavour states ought to be orthogonal \footnote{The situation is similar to the fact that, to this day, right-helicity neutrinos have not been detected, due to the suppression by a factor $m_i^2/\tp^2$ by comparison with their left-helicity counterparts. Nevertheless, the existence of neutrino mass implies their existence.}. Consequently, the possibility of zero-distance conversion among different flavours is not theoretically excluded. 

Since the coherence factors are significant for non-relativistic neutrinos, the present approach will be relevant for the analysis of low-energy neutrino experiments, like the measurement of the absolute mass of the electron neutrino in the KATRIN experiment \cite{KATRIN}, or the planned PTOLEMY \cite{PTOLEMY} experiment for the detection of the Cosmic Neutrino Background. At a conceptual level, our results show that the coherence of oscillating neutrino states can be formulated using a strictly quantum field theoretical scheme, without invoking wave packets and other quantum mechanical notions. This will shed a new light on neutrinos in cosmology \cite{Dolgov}, namely on the analysis of astrophysical neutrino processes, where the coherence and decoherence are essential \cite{Smirnov_decoher, GB_coher, GB_conf, Smirnov_coher}. We shall return to these aspects in further communications.

\subsection*{Acknowledgments}
We are grateful to M. Chaichian and K. Fujikawa for useful comments.

\appendix

\section{Seesaw Hamiltonian diagonalization}\label{diag}

The results of Sect. \ref{osc_seesaw} can be obtained also by the procedure of Hamiltonian diagonalization, without using at all the information provided by the Lagrangian diagonalization briefly reviewed in Sect. \ref{Lagr_formalism}. Below we sketch the steps of this procedure. We use the notations of Sect. \ref{osc_seesaw}.

We start from the Hamiltonian in the form \eqref{H_modes}, which we copy below for facilitating the reading:
\begin{eqnarray}\label{H_modes_app}
H&=&\int d^3 p\ \tp\left(a^\dagger_\da({\bf p}) a_\da({\bf p})+b^\dagger_\ua({\bf p})b_\ua({\bf p})+c^\dagger_\ua({\bf p})c_\ua({\bf p})+d^\dagger_\da({\bf p}) d_\da({\bf p})\right)\cr
&+&i\int d^3 p \Big[m_D\left(a^\dagger_\da({\bf p}) d^\dagger_\da(-{\bf p})+d_\da({\bf p}) a_\da(-{\bf p})-c^\dagger_\ua({\bf p}) b^\dagger_\ua(-{\bf p})-b_\ua({\bf p}) c_\ua(-{\bf p})\right)\cr
&-&\frac{m_L}{2}\left( a^\dagger_\da({\bf p}) a^\dagger_\da(-{\bf p})+a_\da({\bf p}) a_\da(-{\bf p})+b^\dagger_\ua({\bf p}) b^\dagger_\ua(-{\bf p})+b_\ua({\bf p}) b_\ua(-{\bf p}) \right)\cr
&-&\frac{m_R}{2}\left( c^\dagger_\ua({\bf p}) c^\dagger_\ua(-{\bf p})+c_\ua({\bf p}) c_\ua(-{\bf p})+d^\dagger_\da({\bf p}) d^\dagger_\da(-{\bf p})+d_\da({\bf p}) d_\da(-{\bf p}) \right) \Big].
\end{eqnarray}
Without knowing a priori that this Hamiltonian can be diagonalized, we can attempt to achieve the diagonalization by using canonical transformations of the massless particle creation and annihilation operators. The first priority is to isolate the terms with mixed operators, namely those which are proportional to $m_D$ in \eqref{H_modes_app}. This is achieved by a unitary transformation
\begin{eqnarray}\label{rot_abcd_app}
\left(\begin{array}{c}
            a_{1\da}({\bf p})\\
            a_{2\da}({\bf p})
            \end{array}\right)&=&\left(\begin{array}{cc}
            \rho_1^*&0\\
          0&\rho_2^*
            \end{array}\right)\left(\begin{array}{cc}
            \cos\theta&\sin\theta\\
           -\sin\theta&\cos\theta
            \end{array}\right)\left(\begin{array}{c}
            a_{\da}({\bf p})\\
            -d_\da({\bf p})
            \end{array}\right)\cr
\left(\begin{array}{c}
            a_{1\ua}^\dagger({\bf p})\\
           a_{2\ua}^\dagger({\bf p})
            \end{array}\right)&=&\left(\begin{array}{cc}
            \rho_1^*&0\\
          0&\rho_2^*
            \end{array}\right)\left(\begin{array}{cc}
            \cos\theta&-\sin\theta\\
           \sin\theta&\cos\theta
            \end{array}\right)\left(\begin{array}{c}
            b_{\ua}^\dagger({\bf p})\\
            c_\ua^\dagger({\bf p})
            \end{array}\right),
\end{eqnarray}
where the angle $\theta$ is arbitrary and will be later fixed by the requirement of eliminating the mixed products, while $\rho_1$ and $\rho_2$ are phases which will be fixed by the requirement that the physical masses be positive (Autonne--Takagi factorization of a $2\times 2$ unitary matrix).

Let us focus on the left-helicity operators in the Hamiltonian \eqref{H_modes_app} and see what is the effect of this canonical transformation. We have:
\begin{eqnarray}
a_{\da}^\dagger({\bf p})a_{\da}({\bf p})&=&\vert \rho_1 \vert^2 \cos^2\theta\, a_{1\da}^\dagger({\bf p})a_{1\da}({\bf p})+\vert \rho_2 \vert^2 \sin^2\theta\, a_{2\da}^\dagger({\bf p})a_{2\da}({\bf p})\cr
&+&\sin\theta\cos\theta\left(\rho_1^* \rho_2 a_{1\da}^\dagger({\bf p})a_{2\da}({\bf p})+ \rho_1 \rho_2^* a_{2\da}^\dagger({\bf p})a_{1\da}({\bf p})\right),\cr
d_{\da}^\dagger({\bf p})d_{\da}({\bf p})&=&\vert \rho_1 \vert^2 \sin^2\theta\, a_{1\da}^\dagger({\bf p})a_{1\da}({\bf p})+ \vert \rho_2 \vert^2 \cos^2\theta\, a_{2\da}^\dagger({\bf p})a_{2\da}({\bf p})\cr
&-&\sin\theta\cos\theta\left(\rho_1^*\rho_2 a_{1\da}^\dagger({\bf p})a_{2\da}({\bf p})+ \rho_1\rho_2^* a_{2\da}^\dagger({\bf p})a_{1\da}({\bf p})\right),\cr
a_{\da}^\dagger({\bf p})d^\dagger_{\da}(-{\bf p})&=&\sin\theta\cos\theta\, \left([\rho_1^*]^2 a_{1\da}^\dagger({\bf p})a_{1\da}^\dagger(-{\bf p}) - [\rho_2^*]^2 a_{2\da}^\dagger({\bf p})a_{2\da}^\dagger(-{\bf p})\right)\cr
&-& \rho_1^* \rho_2^* \cos^2\theta a_{1\da}^\dagger({\bf p})a_{2\da}\dagger(-{\bf p})+ \rho_1^* \rho_2^* \sin^2\theta\,a_{2\da}^\dagger({\bf p})a_{1\da}^\dagger(-{\bf p}),
\cr
a_{\da}^\dagger({\bf p})a^\dagger_{\da}(-{\bf p})&=&[\rho_1^*]^2 \cos^2\theta\, a_{1\da}^\dagger({\bf p})a_{1\da}^\dagger(-{\bf p}) + [\rho_2^*]^2 \sin^2\theta\, a_{2\da}^\dagger({\bf p})a_{2\da}^\dagger(-{\bf p})\cr
&+&\rho_1^* \rho_2^* \sin\theta\cos\theta \left(a_{1\da}^\dagger({\bf p})a_{2\da}^\dagger(-{\bf p})+a_{2\da}^\dagger({\bf p})a_{1\da}^\dagger(-{\bf p})\right),
\cr
d_{\da}^\dagger({\bf p})d^\dagger_{\da}(-{\bf p})&=&[\rho_1^*]^2 \sin^2\theta\, a_{1\da}^\dagger({\bf p})a_{1\da}^\dagger(-{\bf p})+[\rho_2^*]^2 \cos^2\theta\, a_{2\da}^\dagger({\bf p})a_{2\da}^\dagger(-{\bf p})\cr
&-&\rho_1^* \rho_2^* \sin\theta\cos\theta \left(a_{1\da}^\dagger({\bf p})a_{2\da}^\dagger(-{\bf p})+a_{2\da}^\dagger({\bf p})a_{1\da}^\dagger(-{\bf p})\right).
\end{eqnarray}
As a result, we find that
\begin{eqnarray}
a_{\da}^\dagger({\bf p})a_{\da}({\bf p})+d_{\da}^\dagger({\bf p})d_{\da}({\bf p})=a_{1\da}^\dagger({\bf p})a_{1\da}({\bf p})+a_{2\da}^\dagger({\bf p})a_{2\da}({\bf p}),
\end{eqnarray}
as well as
\begin{eqnarray}\label{calc1}
im_D a_{\da}^\dagger({\bf p})d^\dagger_{\da}(-{\bf p})-i\frac{m_L}{2}a_{\da}^\dagger({\bf p})a^\dagger_{\da}(-{\bf p})-i\frac{m_R}{2}d_{\da}^\dagger({\bf p})d^\dagger_{\da}(-{\bf p})\cr
=i \rho_1^*\rho_2^* a_{1\da}^\dagger({\bf p})a_{2\da}^\dagger(-{\bf p})\left[-m_D\cos^2\theta + \frac{m_R-m_L}{2}\sin\theta\cos\theta\right]\cr
+ i \rho_1^*\rho_2^* a_{2\da}^\dagger({\bf p})a_{1\da}^\dagger(-{\bf p})\left[m_D\sin^2\theta+\frac{m_R-m_L}{2}\sin\theta\cos\theta\right]\cr
+ i [\rho_1^*]^2 a_{1\da}^\dagger({\bf p})a_{1\da}^\dagger(-{\bf p})\left[m_D\sin\theta\cos\theta-\frac{m_L}{2}\cos^2\theta-\frac{m_R}{2}\sin^2\theta\right]\cr
+ i [\rho_2^*]^2 a_{2\da}^\dagger({\bf p})a_{2\da}^\dagger(-{\bf p})\left[-m_D\sin\theta\cos\theta-\frac{m_L}{2}\sin^2\theta-\frac{m_R}{2}\cos^2\theta\right].
\end{eqnarray}
To eliminate the combinations $a_{1\da}^\dagger a_{2\da}^\dagger$, we request the corresponding coefficient in \eqref{calc1} to vanish. Note that $a_{2\da}^\dagger({\bf p})a_{1\da}^\dagger(-{\bf p}) = a_{1\da}^\dagger({\bf p})a_{2\da}^\dagger(-{\bf p})$ due to both the commutation relations of the creation operators as well as the spinor structure of the Hamiltonian \eqref{Hamilt_osc}. Hence,
\begin{eqnarray}
m_D(\sin^2\theta - \cos^2\theta)+\frac{m_R-m_L}{2}\,2\sin\theta\cos\theta=0,
\end{eqnarray}
leading to
\begin{eqnarray}\label{theta'}
\tan 2\theta=\frac{2m_D}{m_R-m_L}.
\end{eqnarray}
Note that we obtained, as expected, the same angle $\theta$ as the one found in the process of Lagrangian diagonalization, eq. \eqref{theta}.

The coefficient of $ia_{1\da}^\dagger({\bf p})a_{1\da}^\dagger(-{\bf p})$ is interpreted as a mass:
\begin{eqnarray}
[\rho_1^*]^2 \Big(m_D\sin\theta\cos\theta-\frac{m_L}{2}\cos^2\theta-\frac{m_R}{2}\sin^2\theta\Big)&=&-\frac{m_1}{2}.
\end{eqnarray}
Same for the coefficient of $ia_{2\da}^\dagger({\bf p})a_{2\da}^\dagger(-{\bf p})$:
\begin{eqnarray}
[\rho_2^*]^2 \Big(-m_D\sin\theta\cos\theta-\frac{m_L}{2}\sin^2\theta-\frac{m_R}{2}\cos^2\theta\Big)&=&-\frac{m_2}{2}.
\end{eqnarray}
In a similar manner, one expresses all the terms in \eqref{H_modes_app} in terms of the operators $a_{1\lambda}, a^\dagger_{1\lambda}$ and $a_{2\lambda}, a^\dagger_{2\lambda}$.

The Hamiltonian, after the unitary transformation \eqref{rot_abcd}, becomes:
\begin{eqnarray}\label{H_modes'}
H&=&\int d^3 p\ \sum_\lambda\Big[\tp\left(a_{1\lambda}^\dagger({\bf p})a_{1\lambda}({\bf p})+a_{2\lambda}^\dagger({\bf p})a_{2\lambda}({\bf p})\right)\cr
&-&i\frac{m_1}{2}\left( a_{1\lambda}^\dagger({\bf p})a_{1\lambda}^\dagger(-{\bf p})+a_{1\lambda}({\bf p})a_{1\lambda}(-{\bf p}) \right)\cr
&-&i\frac{m_2}{2}\left( a_{2\lambda}^\dagger({\bf p})a_{2\lambda}^\dagger(-{\bf p})+a_{2\lambda}({\bf p})a_{2\lambda}(-{\bf p}) \right)\Big]\Big\},
\end{eqnarray}
with real and positive parameters $m_{1,2}$ once the factors $\rho_{1,2}$ have been fixed to fulfill this condition. We note that the Hamiltonian \eqref{H_modes'} has exactly the form of the BCS Hamiltonian in the theory of superconductivity. It is well-known that such a Hamiltonian is diagonalized by Bogoliubov transformations. We will need two sets of such transformations,
\begin{eqnarray}\label{Btrans_app}
A_{1\lambda}({\bf p})&=&\alpha_{1\tp}a_{1\lambda}({\bf p})+i\beta_{1\tp}\,a^\dagger_{1\lambda}(-{\bf p}),\cr
A_{2\lambda}({\bf p})&=&\alpha_{2\tp}a_{2\lambda}({\bf p})+i\beta_{2\tp}\,a^\dagger_{2\lambda}(-{\bf p}),
\end{eqnarray}
where $\alpha_{i\tp},\ \beta_{i\tp}$, $i=1,2$  are complex coefficients to be determined. To find them, we make the ansatz that the Hamiltonian \eqref{H_modes'}, after applying the transformations \eqref{Btrans_app}, becomes
\begin{eqnarray}\label{H_diag_app}
H&=\int d^3 p \sum_{\lambda} \Big[ E_{1\tp} A^\dagger_{1\lambda}({\bf p}) A_{1\lambda}({\bf p})+ E_{2\tp} A^\dagger_{2\lambda}({\bf p}) A_{2\lambda}({\bf p})\Big].
\end{eqnarray}
Plugging in \eqref{H_diag_app} the transformations \eqref{Btrans_app}, we find that the two forms of the Hamiltonian coincide if
\begin{eqnarray}
|\alpha_{i\tp}|^2-|\beta_{i\tp}|^2&=&\frac{\tp}{ E_{i\tp}},\cr
\alpha_{i\tp}\beta_{i\tp}&=&-\frac{m_i}{2 E_{i\tp}},\ \ \ \ i=1,2.
\end{eqnarray}
This system is solved by
\begin{eqnarray}\label{BT_coeff}
\alpha_{i\tp}=\sqrt{\frac{ E_{i\tp}+\tp}{2 E_{i\tp}}},\ \ \ \ \ 
\beta_{i\tp}=\sqrt{\frac{ E_{i\tp}-\tp}{2 E_{i\tp}}},\ \ \ \  E_{i\tp}=\sqrt{m_i^2+\tp^2}.
\end{eqnarray}

In conclusion, the method of Hamiltonian diagonalization presented above gives identical results with the method presented in Sect. \ref{osc_seesaw}. While the idea of diagonalizing the Hamiltonian is physically transparent, it is however technically more tedious. Nevertheless, it justifies the "shortcut" method described in Sect. \ref{osc_seesaw}.

To make an analogy with the dynamical generation of nucleon masses, the procedure for establishing the vacuum structure described in Sect. \ref{osc_seesaw} is analogous with the one presented by Nambu and Jona-Lasinio in \cite{NJL}, while the procedure described in this appendix is analogous to the one detailed by Umezawa, Takahashi and Kamefuchi in \cite{UTK}.

\section{Conventions for spinors}\label{appendix1}

We work with the chiral representation of the $\gamma$-matrices:
\begin{eqnarray}\label{gamma_matr}
\gamma^0=\left(\begin{array}{cc}
            0&\sigma^0\\
            \sigma^0&0
            \end{array}\right),\ \ \ \ 
\gamma^i=\left(\begin{array}{cc}
            0&\sigma^i\\
            -\sigma^i&0
            \end{array}\right),\ \ \ \
\gamma_5=\left(\begin{array}{cc}
            -\sigma^0&0\\
           0&\sigma^0
            \end{array}\right),
\end{eqnarray}
where $\sigma^0={\mathbf 1}_{2\times 2}$ and $\sigma^i$, $i=1,2,3$ are the Pauli matrices. The charge conjugation matrix is
\begin{eqnarray}\label{C_matr}
C=i\gamma^2\gamma^0=i\left(\begin{array}{cc}
            \sigma^2&0\\
            0&-\sigma^2
            \end{array}\right),\ \ \ \ 
\end{eqnarray}
with the properties: $C^T=C^\dagger=-C$.

The solution of the Dirac equation
\begin{eqnarray}\label{D_eq}
(i\gamma^\mu\partial_\mu-m)\psi(x)=0
\end{eqnarray}
is written in mode expansion as
\begin{eqnarray}\label{mode_exp}
\psi(x)=\int\frac{d^3p}{(2\pi)^{3/2}\sqrt{2 E_\tp}}\sum_\lambda\left(a_\lambda({\bf p})U_\lambda(m,{\bf p})e^{-ipx}+b^\dagger_\lambda({\bf p})V_\lambda(m,{\bf p})e^{ipx}\right),
\end{eqnarray}
where $\lambda=\pm{\frac{1}{2}}$ are the helicity eigenvalues and $p_0= E_\tp=\sqrt{\tp^2+m^2}$, with the notation $\text{p}=|{\bf p}|$. The spinors $U_\lambda(m,{\bf p})$ and $V_\lambda(m,{\bf p})$ are helicity eigenvectors,
\begin{eqnarray}
\frac{\hat{\bf S}\cdot{\bf p}}{\tp}U_\lambda(m,{\bf p})=\lambda U_\lambda(m,{\bf p}),\ \ \ \frac{\hat{\bf S}\cdot{\bf p}}{\tp}V_\lambda(m,{\bf p})=-\lambda V_\lambda(m,{\bf p}),
\end{eqnarray}
with the spin matrix
\begin{eqnarray}
\hat{\bf S}^i= \frac{1}{2}{\bf \Sigma}^i=\left(\begin{array}{cc}
            {\sigma}^i&0\\
            0&\sigma^i
            \end{array}\right).
\end{eqnarray}
We also employ the notation
\begin{eqnarray}\label{small_spinors}
u_\lambda({\bf p}) = U_\lambda(0,{\bf p}),
\end{eqnarray}
and similar for $v_\lambda({\bf p})$. The left- and right-handed helicity spinors read:
\begin{eqnarray}\label{helicity_spinors}
&U_{\uparrow}(m,{\bf p})=\sqrt{\frac{ E_\tp+m}{2}}\left(\begin{array}{c}
            \left(1-\frac{\text{p}}{ E_\tp+m}\right)\chi_\uparrow\\
            \left(1+\frac{\text{p}}{ E_\tp+m}\right)\chi_\uparrow
            \end{array}\right),\ \ \ 
U_{\downarrow}(m,{\bf p})=\sqrt{\frac{ E_\tp+m}{2}}\left(\begin{array}{c}
            \left(1+\frac{\text{p}}{ E_\tp+m}\right)\chi_\downarrow\\
            \left(1-\frac{\text{p}}{ E_\tp+m}\right)\chi_\downarrow
            \end{array}\right),\\ 
&V_{\uparrow}(m,{\bf p})=\sqrt{\frac{ E_\tp+m}{2}}\left(\begin{array}{c}
            -\left(1+\frac{\text{p}}{ E_\tp+m}\right)\chi_\downarrow\\
            \left(1-\frac{\text{p}}{ E_\tp+m}\right)\chi_\downarrow
            \end{array}\right),\ \ \ 
V_{\downarrow}(m,{\bf p})=\sqrt{\frac{ E_\tp+m}{2}}\left(\begin{array}{c}
            \left(1-\frac{\text{p}}{ E_\tp+m}\right)\chi_\uparrow\\
            -\left(1+\frac{\text{p}}{ E_\tp+m}\right)\chi_\uparrow
            \end{array}\right)\ ,\nonumber
\end{eqnarray}
and
\begin{eqnarray}\label{helicity_spinors_minus}
&U_{\uparrow}(m,-{\bf p})=i\sqrt{\frac{ E_\tp+m}{2}}\left(\begin{array}{c}
            \left(1-\frac{\text{p}}{ E_\tp+m}\right)\chi_\downarrow\\
            \left(1+\frac{\text{p}}{ E_\tp+m}\right)\chi_\downarrow
            \end{array}\right)\ ,\ \ \ 
V_{\downarrow}(m,-{\bf p})=i\sqrt{\frac{ E_\tp+m}{2}}\left(\begin{array}{c}
            \left(1+\frac{\text{p}}{ E_\tp+m}\right)\chi_\uparrow\\
            \left(1-\frac{\text{p}}{ E_\tp+m}\right)\chi_\uparrow
            \end{array}\right)\ ,\ \ \ \cr
&V_{\uparrow}(m,-{\bf p})=i\sqrt{\frac{ E_\tp+m}{2}}\left(\begin{array}{c}
            -\left(1+\frac{\text{p}}{ E_\tp+m}\right)\chi_\uparrow\\
            \left(1-\frac{\text{p}}{ E_\tp+m}\right)\chi_\uparrow
            \end{array}\right)\ ,\ \ \ 
V_{\downarrow}(m,-{\bf p})=i\sqrt{\frac{ E_\tp+m}{2}}\left(\begin{array}{c}
            \left(1-\frac{\text{p}}{ E_\tp+m}\right)\chi_\downarrow\\
            -\left(1+\frac{\text{p}}{ E_\tp+m}\right)\chi_\downarrow
            \end{array}\right), \nonumber \\
\end{eqnarray}
where the symbol $\uparrow$ denotes the right-handed spinor, while $\downarrow$ denotes the left-handed spinor.
We use the helicity basis 
\begin{eqnarray}\label{helicity_basis}
\chi_{\uparrow}=\eta_\downarrow=\left(\begin{array}{c}
            \cos\frac{\theta}{2}e^{-i\frac{\phi}{2}}\\
             \sin\frac{\theta}{2}e^{i\frac{\phi}{2}}
            \end{array}\right)\ ,\ \ \ 
\chi_{\downarrow}=\eta_\uparrow=\left(\begin{array}{c}
            -\sin\frac{\theta}{2}e^{-i\frac{\phi}{2}}\\
             \cos\frac{\theta}{2}e^{i\frac{\phi}{2}}
            \end{array}\right),
\end{eqnarray}
with $\theta$ and $\phi$ being the polar and azimuthal angles of the momentum vector, \\${\bf p}=(\text{p}\sin\theta\cos\phi,\ \text{p}\sin\theta\sin\phi,\ \text{p}\cos\theta)$. The basis spinors $\chi_\lambda$ and $\eta_\lambda$ satisfy
\begin{eqnarray}
({\vec \sigma}\cdot{\bf p})\chi_\lambda=2\lambda\,\tp\, \chi_\lambda,  \ \ \ \ ({\vec \sigma}\cdot{\bf p})\eta_\lambda=-2\lambda\,\tp\,\eta_\lambda
\end{eqnarray}
and are normalized as
\begin{eqnarray}\label{basis}
\chi^\dagger_\lambda\chi_{\lambda'}=\eta^\dagger_\lambda\eta_{\lambda'}=\delta_{\lambda\lambda'}, \ \rm{where}\ \  \lambda, \lambda'=\pm\frac{1}{2}.
\end{eqnarray}
We note as well the relations:
\begin{eqnarray}
&&\chi_\uparrow\chi^\dagger_\uparrow+\chi_\downarrow\chi^\dagger_\downarrow={\mathbf 1}_{2\times2},\\
&&\chi_\uparrow\chi^\dagger_\uparrow-\chi_\downarrow\chi^\dagger_\downarrow=\left(\begin{array}{cc}
            \cos{\theta}&\sin\theta e^{-i{\phi}}\\
             \sin{\theta}e^{i{\phi}}&-\cos\theta
            \end{array}\right)=\frac{{\vec \sigma}\cdot{\bf p}}{\tp}.
\end{eqnarray}

The helicity spinors  are normalized as
\begin{eqnarray}\label{spinor_norm}
U^\dagger_\lambda(m, {\bf p})U_{\lambda'}(m, {\bf p})&=&2 E_\tp\delta_{\lambda\lambda'},\cr
U^\dagger_\lambda(m, {\bf p})V_{\lambda'}(m, -{\bf p})&=&0,
\end{eqnarray}
and
satisfy the relations:
\begin{eqnarray}\label{spinor_rel}
\bar U_\lambda(m, {\bf p})U_{\lambda'}(m, {\bf p})&=&2m\delta_{\lambda\lambda'},\cr
\bar V_\lambda(m, {\bf p})V_{\lambda'}(m, {\bf p})&=&-2m\delta_{\lambda\lambda'},\cr
\bar U_\lambda(m, {\bf p})V_{\lambda'}(m, -{\bf p})&=&-2i{\text p}\,{\text {sgn}}\lambda\, \delta_{\lambda\lambda'},\cr
\bar V_\lambda(m, {\bf p})U_{\lambda'}(m, -{\bf p})&=&-2i{\text p}\,{\text {sgn}}\lambda\, \delta_{\lambda\lambda'},
\end{eqnarray}
as well as
\begin{eqnarray}\label{spinor_rel_5}
\bar U_\lambda(m, {\bf p})\gamma_5 U_{\lambda'}(m, {\bf p})&=&0,\cr
\bar V_\lambda(m, {\bf p})\gamma_5 V_{\lambda'}(m, {\bf p})&=&0,\cr
\bar U_\lambda(m, {\bf p})\gamma_5 V_{\lambda'}(m, -{\bf p})&=&2i E_\tp \delta_{\lambda\lambda'},\cr
\bar V_\lambda(m, {\bf p})\gamma_5 U_{\lambda'}(m, -{\bf p})&=&-2i E_\tp \delta_{\lambda\lambda'}.
\end{eqnarray}
%

%%%%%%%%%%%%%%%%%%%%%%%%%%%%%%%%%%%%%5

\end{document}